\documentstyle[epsfig]{mn}
\begin{document}

\title[]
{Powerful jets from black hole X-ray binaries in Low/Hard X-ray states}
\author[R.~P.~Fender]
{R. P. Fender\thanks{email : rpf@astro.uva.nl}\\
Astronomical Institute `Anton Pannekoek', University of Amsterdam,
and Center for High Energy Astrophysics, Kruislaan 403, \\
1098 SJ, Amsterdam, The Netherlands\\
}

\maketitle

\begin{abstract}

Four persistent (Cygnus X-1, GX 339-4, GRS 1758-258 and 1E 1740.7-2942)
and three transient (GS 2023+38, GRO J0422+32 and GS 1354-64) black
hole X-ray binary systems have been extensively observed at radio
wavelengths during extended periods in the Low/Hard X-ray state, which
is characterised in X-rays by a hard power-law spectrum and strong
variability. All seven systems show a persistent flat or inverted (in
the sense that $\alpha \ga 0$, where $S_{\nu} \propto \nu^{\alpha}$)
radio spectrum in this state, markedly different from the optically
thin radio spectra exhibited by most X-ray transients within days of
outburst. Furthermore, in none of the systems is a high-frequency
cut-off to this spectral component detected, and there is evidence
that it extends to near-infrared or optical regimes. Luminous
persistent hard X-ray states in the black hole system GRS 1915+105
produce a comparable spectrum. This spectral component is considered
to arise in synchrotron emission from a conical, partially
self-absorbed jet, of the same genre as those originally considered
for Active Galactic Nuclei.  Whatever the physical origin of the
Low/Hard X-ray states, these self-similar outflows are an ever-present
feature.  The power in the jet component is likely to be a significant
($\geq 5$\%) and approximately fixed fraction of the total accretion
luminosity. The correlation between hard X-ray and synchrotron
emission in all the sources implies that the jets are intimately
related to the Comptonisation process, and do not have very large bulk
Lorentz factors, unless the hard X-ray emission is also beamed by the
same factor.

\end{abstract}
\begin{keywords}

binaries: close -- ISM:jets and outflows -- radio continuum: stars
-- X-rays: stars

\end{keywords}


\section{Introduction} 

Radio emission is often observed from X-ray binaries, particularly
transient systems, and especially the black hole candidates. It is
increasingly accepted that this radio emission is the radiative
signature of jet-like outflows, some or all of which may possess
relativistic bulk motions. Recent reviews may be found in Hjellming \&
Han (1995); Mirabel \& Rodr\'\i guez (1999) and Fender (2000). 

\subsection{Black Hole X-ray states}

The `Low/Hard' X-ray state is one of five `canonical' X-ray states,
characterised by both spectral and timing behaviour at X-ray energies,
observed from black hole X-ray binaries in our Galaxy (e.g. Tanaka \&
Lewin 1995; Nowak 1995; Mendez, Belloni \& van der Klis 1998; Grove et
al. 1998; Poutanen 1998).  As discussed in Fender (2000), the relation
of black hole X-ray state to radio emission can be summarised as in Table 1.
In addition, transitions between states appear to produce discrete
ejection events in both persistent and transient (where rapid state
changes are observed as outbursts) systems (Hjellming \& Han 1995;
Kuulkers et al. 1999; Fender et al. 1999a,b, Fender \& Kuulkers 2001).

The X-ray spectrum in the Low/Hard state is dominated by a power-law
component which extends to $\ga 100$ keV (Poutanen 1998 and references
therein; but see also McConnell et al. 2000);
any thermal (accretion disc) component contributes $\la 20$\% in the
soft X-ray band ($\la 10$ keV) and not at all in hard ($\ga 10$ keV)
X-rays.  Modelling of the Low/Hard state favours an accretion disc
which is truncated at some distance (typically inferred to be of order
100 Schwarzschild radii) from the central black hole, around which
exists a Comptonising corona (Poutanen 1998).  `Seed' photons
Compton-upscattered 
by the corona produce the observed power-law spectral component; the
apparent cut-off in the power-law component at $\sim 100$ keV is
generally interpreted as evidence for a {\em thermal} distribution of hot
electrons in the corona.  The flow from the inner edge of the
accretion disc appears to be radiatively inefficient and has been
modelled as a two-temperature
advection dominated accretion flow (ADAF; Narayan, Mahadevan \&
Quataert 1998 and references therein).

\section{Persistent X-ray binaries in the Low/Hard state}

The persistent black hole X-ray binaries displaying the Low/Hard X-ray
state are Cyg X-1, GX 339-4 and probably the Galactic Centre
hard-X-ray sources 1E1740.7-2942 \& GRS 1758-258 (e.g. Tanaka \& Lewin
1995). LMC X-3 may also make transient, quasi-periodic excursions from
the High/Soft to the Low/Hard X-ray state (Wilms et al. 2001), but is
too distant for current radio telescopes to observe easily.  We
describe below the radio properties of these four sources; radio spectra
and images are presented in Figs 1 \& 2, additional information on the
radio spectrum, optical flux and distance and presented in table 2.

Note that throughout this paper we refer to the spectral index,
defined as $\alpha = \Delta \log S_{\nu} / \Delta \log \nu$, i.e.
$S_{\nu} \propto \nu^{\alpha}$. Sources with $\alpha \sim 0$ will be
refer to as `flat spectrum', those with $\alpha > 0$ as having
`inverted' spectra. Optically thin emission typically has $-1 \leq
\alpha \leq -0.5$.

\begin{table*}
\centering
\caption{The relation of radio emission to black hole X-ray
state. VHS, IS, HS, LS and PL correspond to Very High State,
Intermediate state, High (soft) state, Low (hard) state and Power Law,
respectively.} 
\begin{tabular}{cccc}
\hline
State & \multicolumn{2}{c}{X-ray properties} & Radio properties \\
      & Spectra & Timing & \\
\hline
VHS / IS & Disc + PL in varying proportions &
5--15\% rms variability, QPOs & ?  \\
HS & Disc ($kT \sim 1$ keV) + weak PL ($\alpha \sim
-1.5$) & $<10$\% rms variability & Radio suppressed by factor $\geq 25$ \\
LS / Off & PL ($\alpha \sim -0.5$) dominated & Up to 50\% rms below
$\sim 1$ Hz & Low level, steady, flat spectrum \\
\hline
\end{tabular}
\end{table*}

\begin{figure*}
\centerline{\epsfig{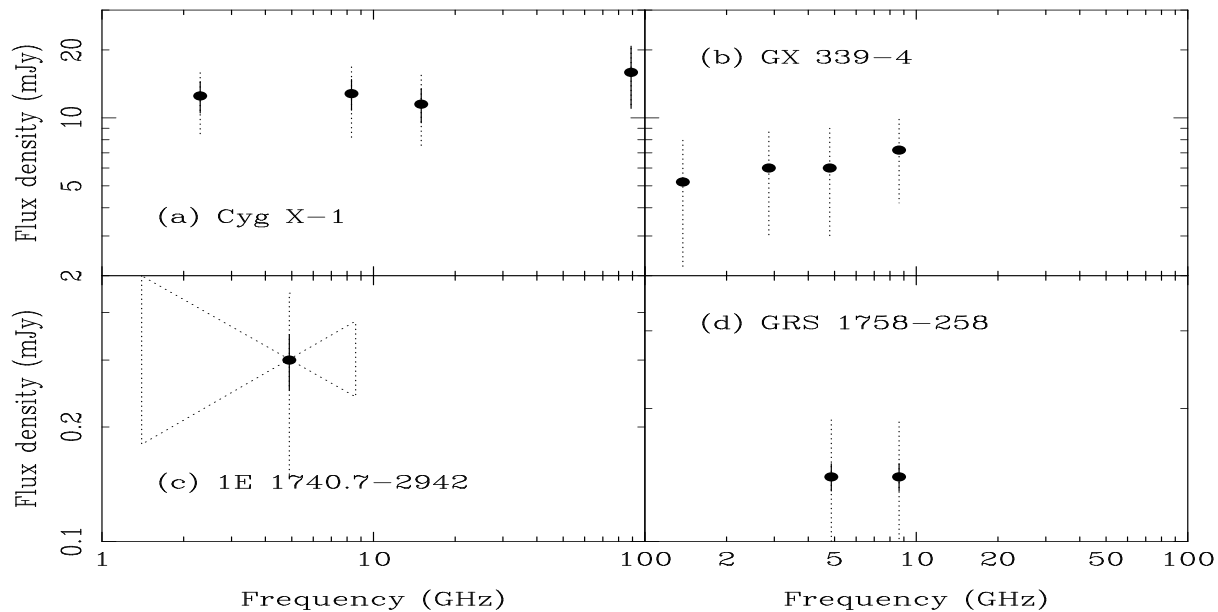}}
\caption{The flat radio(--mm) spectra of the four persistent Galactic
black hole candidate X-ray binaries in the Low/Hard X-ray state.
Solid bars indicate typical measurement uncertainties, dotted bars are
estimates of the observed range of variability in the Low/Hard state.
For 1E 1740.7-2942 the only repeated detections have been at 4.9 GHz;
the range of inferred spectral indices from 1 -- 9 GHz, from a limited
number of observations have been indicated. See main text for references.}
\end{figure*}


\begin{figure*}
\centerline
{\epsfig{file=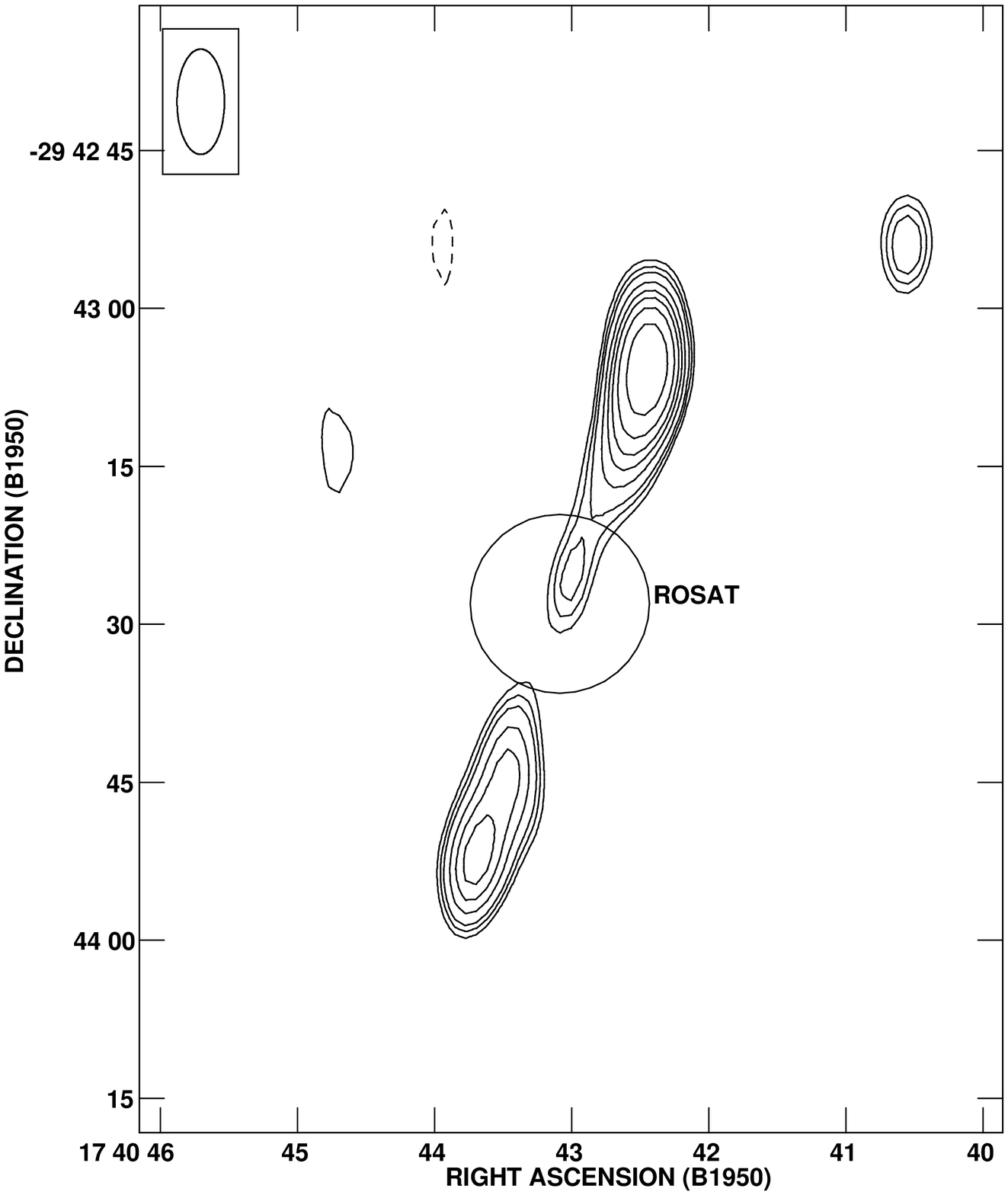, height=8.5cm}\quad
\epsfig{file=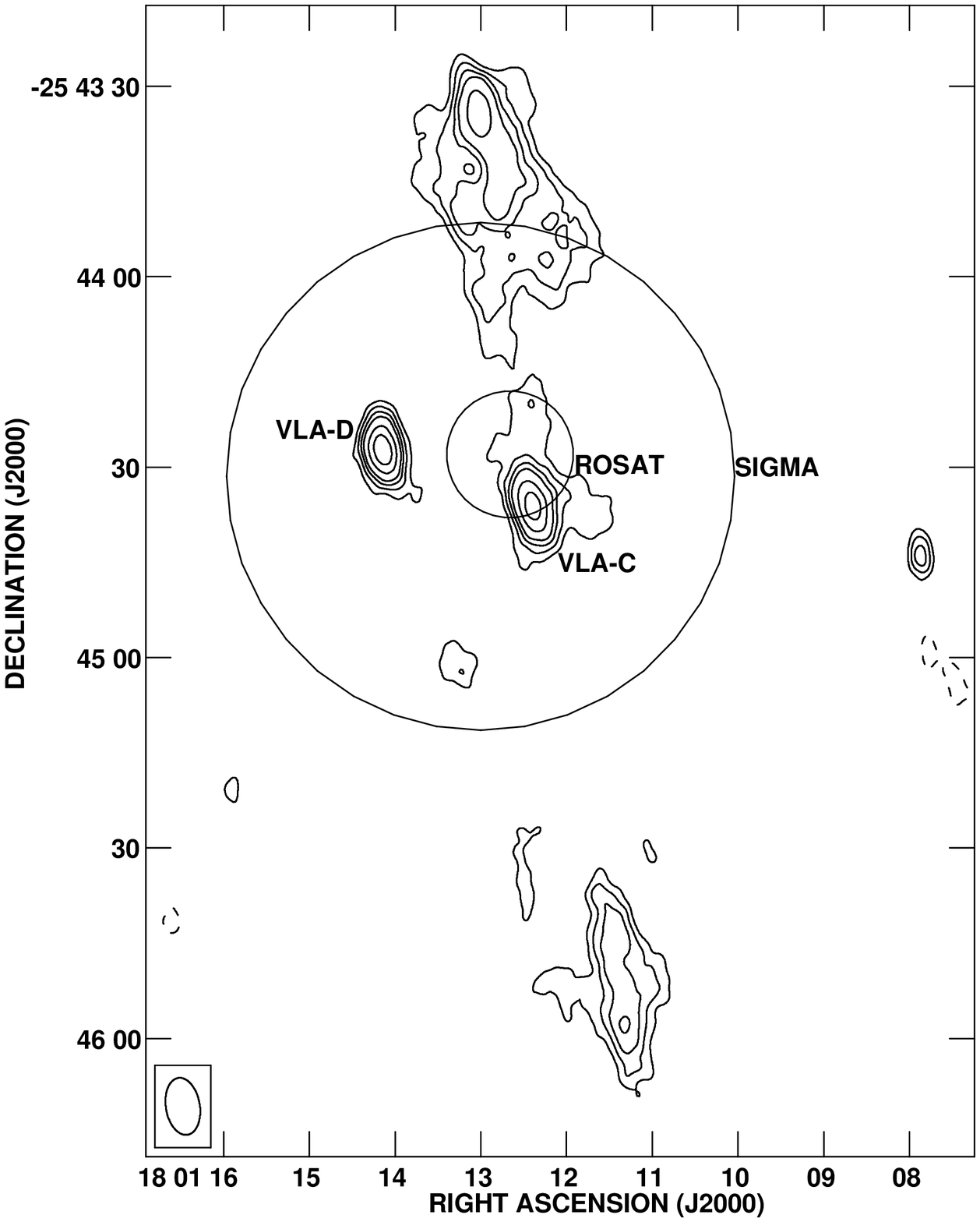, height=8.5cm}}
\caption{Direct imaging of
jets from persistent black hole systems in the Low/Hard X-ray state.
The panels show
arcmin-scale radio jets observed from the two Galactic centre
systems 1E1740.7-2942 (left, Mirabel \& Rodr\'\i guez 1999) and GRS 1758-258 (right, 
Mart\'\i{} et al. 1998). At the (presumed) distance of the
galactic centre, these correspond to physical structures on parsec scales.}
\end{figure*}

\subsection{Cyg X-1}

The radio flux from Cyg X-1 has been established to show a moderate
degree of variability, with a mean flux density of $\sim 15$ mJy at cm
wavelengths (Brocksopp et al. 1999 and references therein). The radio
spectrum extends to mm wavelengths with a spectral index $\alpha \sim
0$ (Fender et al. 2000a). The radio emission is also modulated with a
semi-amplitude of around 20\%, at the 5.6-day orbital period (Pooley,
Fender \& Brocksopp 1999), probably as a result of free-free
absorption in the dense stellar wind of the OB-type companion star
(Brocksopp et al. 1999). On longer (weeks to months) timescales the
radio emission is correlated with soft- and hard-X-ray emission when
the source is in the Low/Hard state (Brocksopp et al. 1999).  There is
strong but inconclusive evidence that during transitions to softer
X-ray states the radio emission is suppressed, and flaring may occur
around the time of state transition (Brocksopp et al. 1999 and
references therein). Cyg X-1 probably lies at a distance of 2--3 kpc
(Fender \& Hendry 2000 and references therein).  Fig 1(a) displays the
mean radio--mm spectrum of Cyg X-1.  Stirling et al. (1998, 2001)
present evidence for an extended and collimated radio structure from
Cyg X-1 on milliarcsecond scales, which is almost certainly a compact
jet.

\subsection{GX 339-4}

GX 339-4 displays a slightly inverted ($\alpha \sim +0.1$) radio
spectrum when in the Low/Hard (and maybe also `Off') X-ray state
(Corbel et al. 2000 and references therein), which may extend to
near-infrared wavelengths (Corbel \& Fender 2000). As in Cyg X-1, the
radio emission is correlated with the soft- and hard-X-ray emission when in
the Low/Hard X-ray state (Hannikainen et al. 1998b; Corbel et
al. 2000) and probably also in the `Off' state (Corbel et al. 2000).
In the High/Soft X-ray state both the hard X-ray and radio emission
from GX 339-4 are strongly suppressed; the source was undetectable
at radio wavelengths throughout 1998 whilst in this state (Fender et
al. 1999b). GX 339-4 probably lies at a distance of 2--4 kpc (Fender \&
Hendry 2000 and references therein).
Fig 1(b) displays a typical radio spectrum for GX 339-4.

\subsection{1E 1740.7-2942}

The radio counterpart to the galactic centre hard X-ray source 1E
1740.7-2942 (Tanaka \& Lewin 1995; also known as the `Great
Annihilator') consists of a compact core at the centre of extended
double-sided jets (Mirabel et al. 1992; Mirabel 1994).  The core radio
emission appears to be correlated with X-ray emission (Mirabel et
al. 1992, 1993). A small number of reported radio observations
(e.g. Mirabel et al. 1992, 1993; Gray, Cram \& Ekers 1992;
Anantharamaiah et al. 1993; Mart\'\i{} 1993) repeatedly detect the core at
5 GHz, but a good multiple-frequency spectrum has yet to be
reported. Several observations report two-point spectral indices, for
example Mirabel et al. (1992, 1993) report 20--6 cm spectral indices
of $\alpha > -0.2$ in 1989 March, and $\alpha = -0.4 \pm 0.1$ in 1991
October. Gray et al. (1992) similarly conclude that the 20--6 cm
spectral index of the core component (their source `A') must be
flatter than -0.36, based on VLA and ATCA observations in 1989--1991.
Heindl, Prince \& Grunsfeld (1994) further report that in 1989 March
while the source was $\sim 0.4$ mJy at 5 GHz it was `undetected' at
1.5 GHz, again implying a fairly flat or inverted spectrum.
However, it is clear that for this source a well determined
multi-frequency radio spectrum is desirable and, at present, lacking.
In Fig 1(c) we indicate the mean 5 GHz flux density and the possible
range of 1--9 GHz spectral indices inferred from the
literature. Certainly the core does not have the steep, optically thin
spectral index of $-0.8\pm 0.1$ observed from the jet lobes/hotspots
(Mirabel et al. 1992).  Fig 2 shows a deep radio image of the core +
jets radio structure of the system (from Mirabel \& Rodr\'\i guez 1999).
1E 1740.7-2942 is assumed to lie at a distance of $\sim 8.5$ kpc, in
the region of the galactic centre (e.g. Tanaka \& Lewin 1995).

\subsection{GRS 1758-258}

GRS 1758-258 is another galactic centre hard X-ray source with very
similar X-ray properties to 1E 1740.7-2942 (Tanaka \& Lewin 1995; Lin
et al. 2000). The radio counterpart is also a compact core with
extended jet-like structures (Mirabel 1994). The compact core has a
flat radio spectrum at cm wavelengths (Lin et al. 2000). Fig 1(d)
shows a typical radio spectrum of the source; Fig 2 shows a deep radio
image of the core + jets radio structure of GRS 1758-258 (from Mart\'\i{}
et al. 1998). Like 1E 1740.7-2942, GRS 1758-258 is also assumed to lie
in the galactic centre region (e.g. Tanaka \& Lewin 1995) and so we
assume a distance of $\sim 8.5$ kpc.

\begin{table*}
\caption{Observed and inferred properties of persistent and transient
BHC X-ray binaries during extended periods in the Low/Hard X-ray state.
Optical flux densities correspond to quasi-simultaneous observations,
and are dereddened. Details of references are provided in the main
text. Note 1 : Cyg X-1 is a high-mass X-ray binary, in which the
optical light is dominated by the OB companion star (all the other
sources are low-mass X-ray binaries).}
\begin{tabular}{rccccccc}
\hline
Source & Epoch & \multicolumn{4}{c}{Radio -- millimetre synchrotron spectrum}
& Optical flux & Distance \\ 
      & & $\nu_{\rm LOW}$ (GHz) & $\nu_{\rm HIGH}$ (GHz) & $\alpha
\pm \sigma$ & $S_{\nu{\rm, HIGH}}$ (mJy) & 
density (mJy) & (kpc)  \\ 
\hline
Cyg X-1 & 1996-1999 & $\sim 1$ & 220 & $0.0 \pm 0.05$ & $\sim 15$ & $\geq 1000^1$ &
$2.0 \pm 0.5$\\
GX 339-4 & 1992-1999 & 0.84 & 8.6$^{1}$ & $0.16 \pm 0.03$ & $\sim 10$ &$\sim 75$ &  $3.0 \pm
1.0$ \\
1E 1740.7-2942 & 1992-1999 &&& -- & $\sim 0.3$ & -- & $8.5 \pm 1.5$\\
GRS 1758-258 & Aug 1997 & 4.8 & 8.6 & $0.0 \pm 0.1$ & $\sim 0.15$ & -- & $8.5 \pm 1.5$\\
\hline
GS 2023+338 & 1989-1991 & 1.49 & 14.9 & $\sim 0 \rightarrow +0.6 $ & $\sim
50$& $\sim 50$ & $\sim 3.5$\\
GRO J0422+32 & 1992 & 1.49 & 14.9 & $\sim 0 \rightarrow +0.1$ & $\sim 2$ &
$\sim 5$ & $\sim 2.2$ \\
GS 1354-64 & 1997 & 0.84 & 8.64 & $\sim 0 \rightarrow +0.2$ & $\sim 2$ &
$\sim 2.5$ & $\sim 10$\\
\hline
\end{tabular}
\end{table*}

\section{X-ray transients in the Low/Hard state}

Only three X-ray transients have been observed to spend an extended
period in the Low/Hard X-ray state during which time there was good
coverage at radio wavelengths. Some systems, such as GRO J1716-249
(`Nova Oph 1993'), {\em were} observed at radio wavelengths during
(probable) excursions in the Low/Hard state, but in these cases the
state changes were so rapid and frequent that the radio emission was
dominated by the major, discrete ejections and the presence (or not)
of an underlying flat spectral component could not be tested.
Distance estimates for these three systems are from Chen et al. (1997).

\begin{figure*}
\centerline{\epsfig{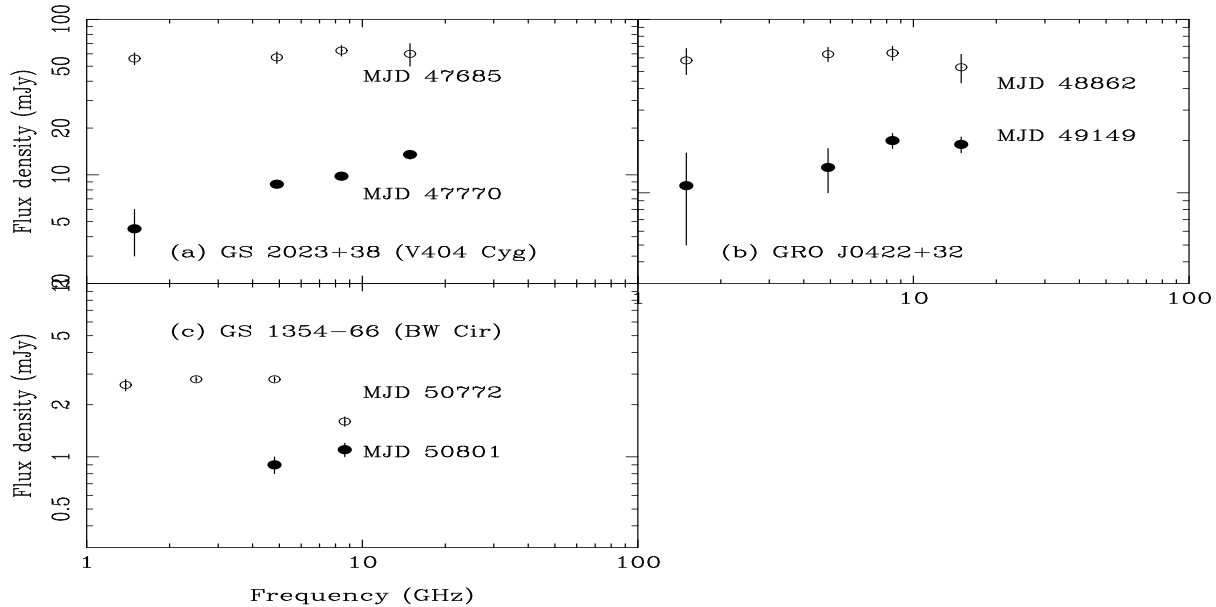}}
\caption{Flat/inverted (i.e. spectral index $\alpha \geq 0$)
radio spectra from three black hole X-ray transients
during an extended period in the Low/Hard X-ray state. Note that all
three sources appear to evolve from approximately flat to inverted
radio spectrum on timescales of tens to hundreds of days.}
\end{figure*}

\subsection{GS 2023+338 (V404 Cyg)}

GS 2023+338 was first detected by the Ginga X-ray satellite in 1989
May (Makino 1989), and was a bright and unusual X-ray transient.
Oosterbroek et al. (1997), in a study of the X-ray spectral and timing
properties of this system, conclude that it stayed in the Low/Hard
X-ray state throughout the outburst (see also Zycki, Done \& Smith
1999a,b).  The radio emission from this system was also unusual -- a
bright optically thin outburst (presumably associated with the rapid
state transition at the start of the outburst) was followed by the
emergence of `second stage' emission with a flat/inverted radio
spectrum (Han \& Hjellming 1992; see Figs 3 and 5).

Han \& Hjellming (1992) also report (amongst other things) rapid
variability, possibly quasi-periodic, which appears, in hindsight, to
be very much like the `radio QPO' observed from GRS 1915+105 (Pooley
\& Fender 1997).  Furthermore, the extremely well correlated radio :
optical light curve (Han \& Hjellming 1992) may be indicating the
extension of the flat spectral component to optical wavelengths (just
as the flat spectral component in GRS 1915+105 extends to at least
near-infrared wavelengths; Fender \& Pooley 1998, 2000 and references
therein). Radio spectra of GS 2023+338 during the Low/Hard state
is shown in Fig 3(a).

\subsection{GRO J0422+32}

GRO J0422+32 was discovered as a new, bright hard X-ray source by
BATSE onboard CGRO in 1992 (Paciesas et al. 1992).  According to Esin
et al. (1998) GRO J0422+32 was at all times during its outburst in the
low/hard X-ray state. Shrader et al. (1994) present the UV, optical
and radio outburst data on this source. The radio observations reveal
a flat radio spectrum which becomes more inverted as the outburst
progresses. Shrader et al. (1994) and Hjellming \& Han (1995) refer to
this flat/inverted radio spectral component as `second stage' radio
emission, and note its similarity to that of GS 2023+338 (see above).
Radio spectra of GRO J0422+32 in the Low/Hard X-ray state is
presented in Fig 3(b).

\subsection{GS 1354-64}

This (recurrent) X-ray transient was discovered by Ginga in 1987
February (Makino 1987). However, it was the 1997 outburst which had
the best coverage, and established that the system remained in the
Low/Hard X-ray state (Revnivtsev et al. 2000; Brocksopp et
al. 2001). During this outburst a radio counterpart to the system was
detected for the first time (Fender et al. 1997c; Brocksopp et
al. 2001).  The radio spectrum of this
system was observed to invert (ie. $\alpha > 0$) within 30 days of the 
initial detection, as the source persisted in the Low/Hard X-ray state
(Fig 3(c); Brocksopp et al. 2001).

\subsection{Transients in the Low/Hard state without good radio coverage}

Miyamoto et al. (1993) show that during the decay phase of its
outburst, the transient GS 1124-683 (`Nova Mus 1991') began a
transition from the high/soft to low/hard X-ray states commencing
around 1991 April 19 (MJD 48365) and being complete by 1991 June 13
(MJD 48420). Unfortunately there appear to be no radio observations
during the Low/Hard state. For a discussion of the optically thin
radio outbursts associated with the very high X-ray state in this
source (Miyamoto et al. 1993), see Ball et al. (1995) and Kuulkers et
al. (1999).

Esin et al. (1998) suggest that the transient GRO J1716-249 was in the
low/hard X-ray state when bright in hard X-rays (as observed with
BATSE - Hjellming et al. 1996). However, the relatively sparse radio
coverage and rapid, repeated state changes (Hjellming et al. 1996),
each one presumably leading to a discrete ejection which evolved to an
optically thin radio state, preclude us from looking for a flat
spectral component associated with the low/hard X-ray state.

Mendez et al. (1998) show that GRO J1655-40 passed through the
Low/Hard X-ray state on its way to `quiescence' ($\equiv$ the `Off'
state ?) at the end of its 1997 outburst. However, as this outburst
had not shown the spectacular radio flaring associated with the
1994/1995 outburst(s), there was no radio coverage being undertaken at
this time. Note that in a detailed study of the radio emission from
GRO J1655-40 during the earlier outburst, Hannikainen et al. (2000) discern the presence of 
underlying flat-spectrum `core' emission which is not highly polarised.

\section{Other X-ray transients}

Generally speaking, most X-ray transients do {\em not} spend any
extended period in the Low/Hard state during outburst, often
transitting from quiescence/Off state to the High/Soft or Very High
states and then fading rapidly away. Reviews of X-ray transient
observational properties can be found in e.g. Tanaka \& Lewin (1995)
and Chen, Shrader \& Livio (1997). Of the 26 transients discussed in
Chen et al. (1997), probably only GRS 2023+338 and GRO J0422+32
remained in the Low/Hard state through (most of) their outbursts.
Esin et al. (2000) discuss the `typical' outbursts of the bright
transients GRS 1124-683 and A0620-00, illustrating the rapid transition
to the Very High state and then the decline through the High,
Intermediate and Low/Hard states back to quiescence ($\equiv$ `Off'
state). The jet source GRO J1655-40 was similarly observed to pass
through the `canonical' X-ray states during the decline from outburst
peak (Mendez et al. 1998).  In such cases of outbursts dominated by
soft X-ray states, the radio emission is often dominated by one or
more discrete ejection events, associated with the state transition
and/or the Very High state. A detailed study of three systems is
presented in Kuulkers et al. (1999 -- note that the data in that paper
on GS 1124-683 correspond to epochs {\em before} the Low/Hard state
was entered); see also Hjellming \& Han (1995). The radio spectra of
these systems, dominated as it is by one or more discrete ejections,
is typically optically thin (i.e. $\alpha \leq -0.5$) within a few
days of the peak of the emission, sometimes temporarily inverting as a
new component (which is initially optically thick) is ejected
(e.g. Kuulkers et al. 1999).  This is as a result of expansion of the
ejecta to the point where self-absorption is no longer important (note
that some transients show an optically thin rise also -- see
e.g. Hjellming et al. 1999 and references therein).  These transient,
optically thin outbursts are often much brighter at radio wavelengths
than the flat spectral components under discussion here, but little is
known about any high-frequency component.

It should be noted that the radio spectra of more unusual radio-bright
X-ray sources such as Cyg X-3 and GRS 1915+105 also evolve towards an
optically thin state following major outbursts (e.g. Hjellming \& Han
1995; Fender et al. 1997a; Fender et al. 1999a), again presumably as
the (discrete) ejecta expand sufficiently for internal self-absorption
to become unimportant.

Fig 4 compares a sample of optically thin radio spectra from five
X-ray transients (for further references see e.g. Hjellming \& Han
1995, Fender \& Kuulkers 2001) plus the unusual radio-jet X-ray
binaries SS 433 and Cyg X-3 (data from Fender et al. 2000b and Fender
et al. 1997b respectively for these two systems) with the seven
flat-spectrum sources presented in Figs 1 and 4. Outburst dates,
distance estimates and associated references are given in table 3; for
XTE J1748-288 we assume a distance of 8.5 kpc given its proximity to
the galactic centre.

While this is not a completely comprehensive sample of data for
optically thin events, it is probably representative of the range of
luminosities observed in optically thin events (Cyg X-3 being the
brightest radio source associated with an X-ray binary -- see
e.g. Fender \& Kuulkers 2001).

Two things are immediately apparent :

\begin{enumerate}
\item{Spectra : the Low/Hard state sources all have flat or inverted
radio(--mm) spectra, whereas the soft transients have optically thin
spectra, most obviously at high frequencies}
\item{Radio luminosity : the optically thin sources span a much larger
range in radio luminosities than the Low/Hard state sources}
\end{enumerate}

\begin{table}
\caption{A sample of optically thin radio outbursts from X-ray
binaries. These are compared with the Low/Hard state sources in Fig 6.}
\center
\begin{tabular}{rccl}
Source & Outburst & Distance & Ref \\
       & Date     & (kpc) \\
\hline
SS 433 & 1999 May & 3 & D98 \\
XTE J1748-288 & 1998 Aug & 8.5 & -- \\
CI Cam & 1998 May & 2 & B99 \\
GRO J1716-249 & 1995 March & 2.4 & C97 \\
GRO J1655-40 & 1994 Aug & 3.2 & C97 \\
Cyg X-3 & 1994 Feb & 8.5 & D83 \\
GS 1124-683 & 1991 Feb & 5.5 & C97 \\
\hline
\end{tabular}
\flushleft
{\bf Refs:} 
D83 -- Dickey (1983),
C97 -- Chen, Shrader \& Livio (1997),
D98 -- Dubner et al. (1998),
B99 -- Belloni et al. (1999)
\end{table}

\begin{figure*}
\leavevmode\epsfig{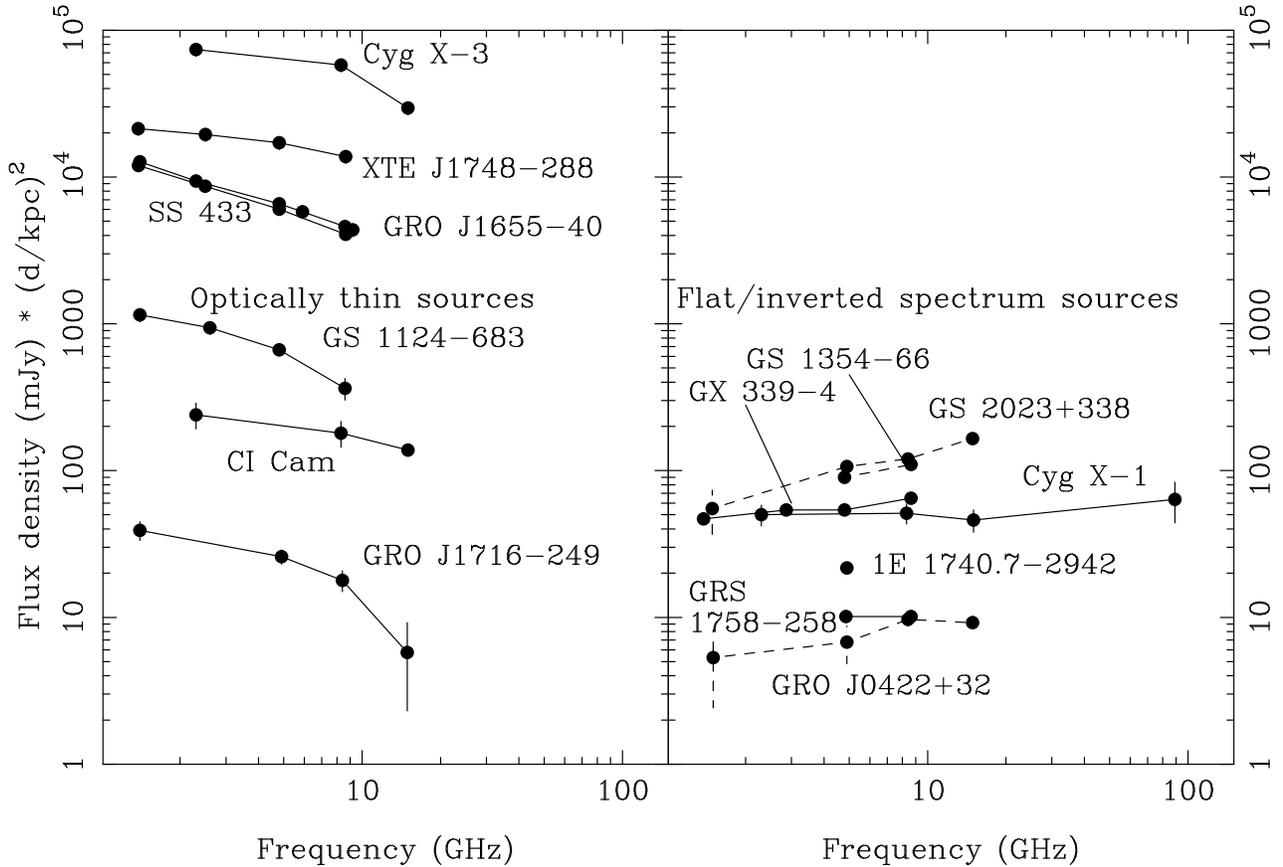}
\caption{Optically thin (i.e. spectral index $\alpha < 0$) radio
spectra from several radio-bright X-ray binaries which were {\em not}
in the Low/Hard X-ray state at the time of the observations, compared
with the flat/inverted spectra of the seven sources in Figs 1 \& 4
(for the transients, the later, ie. most inverted, spectra are
plotted). As well as the different spectral indices (the optically
thin sources all have $-1 \leq \alpha \leq -0.2$, the source in the
Low/Hard state all have $0.0 \leq \alpha \leq 0.6$), note also the
much wider range of fluxes observed from optically thin emission. }
\end{figure*}

\section{Discussion : observational characteristics}

It has been 
shown that, as well as the persistent systems Cyg X-1 and GX
339-4, and also probably 1E 1740.7-2942 and GRS 1758-258, X-ray
transient black hole candidates also show a low-level, flat/inverted
spectral component at radio wavelengths when in the Low/Hard X-ray
state for any length of time.
In this section we discuss the spectral form and extent,
polarisation properties and degree of variability of the flat/inverted
spectral components.

\subsection{Spectral extent and evolution}

As discussed in Fender et al. (2000) for the case of Cyg X-1, as yet
no-one has found either high- or low-frequency cutoffs to the flat
spectral component and as a result the energy associated with it is
essentially unconstrained. However in Cyg X-1 the flat spectral
component is overwhelmed by thermal emission from the OB-type
companion star at wavelengths $\lambda \leq 30\mu$m (Fender et
al. 2000).  In the case of GRS 1915+105 (Fender \& Pooley 1998,2000 and
references therein), and possibly Cyg X-3 (Fender et al. 1996) and GX
339-4 (Corbel \& Fender 2000) there is strong evidence that the flat
spectral component extends to the near-infrared (K-band, $2.2 \mu$m),
and hence has a much larger radiative luminosity ($\geq 10^{36}$ erg
s$^{-1}$ for GRS 1915+105) than could be anticipated from radio
observations only.  However, only one of these systems (GX 339-4) is
in the canonical `Low/Hard' X-ray state (the X-ray states of Cyg X-3
and GRS 1915+105 evade simple classification, although GRS 1915+105
may spend much of its time in something like the Very High State --
Belloni 1998). Do the flat spectral components associated with the
transients in the Low/Hard state also extend to the near-infrared or
beyond ?


Han \& Hjellming (1992) report a very clear correlation between X-ray,
optical and radio fluxes from GS 2023+38 during the decay following
the outburst. The spectral evolution from radio--optical is
illustrated in Fig 5. We assume an extinction in the optical R-band of
2.3 magnitudes (based on $A_{\rm V}=3$ mag -- Shahbaz et al. 1994;
Rieke \& Lebofsky 1985).  The radio--optical spectrum can be fitted by
a single power--law shortly after the emergence of the flat spectral
component around MJD 47685. After that the radio spectrum inverts
while the highest frequency (usually 14.9 GHz) radio flux densities
closely track the dereddened optical flux. Such a strong correlation
implies a common emission mechanism, and we suggest that in this case
the optical flux during this phase of the decay was dominated by
high-frequency synchrotron emission. 

Why the radio spectrum should invert, without affecting the optical
flux, if it is also synchrotron, may be due to free-free absorption
from debris local to the system following the outburst (Zycki, Done \&
Smith 1999a,b do report a large and variable absorption component
present in X-ray spectra following the outburst).  In Appendix A it is
shown that a simple model in which the intrinsic jet emission suffers
foreground free-free absorption can be used to fit the data, but this
is certainly an oversimplification of the true picture. Importantly,
we cannot rule out varying internal synchrotron self-absorption, but
evaluation of its significance would require modelling of the jet,
which is beyond the scope of this work.

The radio spectra of GRO J0422+32 and GS 1354-64 are also observed to
invert as the outburst declines (Fig 3),
in a manner consistent with increasing low-frequency
absorption.  There is also evidence in both cases for an extension of
the flat spectral component to the optical bands.  Van Paradijs et
al. (1994) discuss a flat spectral component from $10 \mu$m through to
the optical band from GRO J0422+32, which they attribute to free-free
emission, most probably from a disc-wind. Shrader et al. (1994) also
note that the dereddened optical continuum of GRO J0422+32 rapidly
evolves to a flat spectrum ($\alpha \sim 0$) during the decay phase of
the outburst.  Finally, Brocksopp et al. (2001) also report evidence
for correlated radio : optical emission from GS 1354-64.

In all three sources the spectrum from the highest radio frequency to
the optical band is approximately flat (table 2). There seems to be no
{\em a priori} reason why flux densities at radio and optical
wavelengths should be comparable in the Low/Hard state, so there
appears to be either a physical coupling, or coincidence.

Very few low-frequency ($\nu < 1$ GHz) observations have been made of
these systems, although in the case of GS 1354-64 the spectrum extends
at least as low as 843 MHz (Brocksopp et al. 2001), as it does in GX 339-4 (Corbel
et al. 2000 and references therein).

\begin{figure}
\centerline{\epsfig{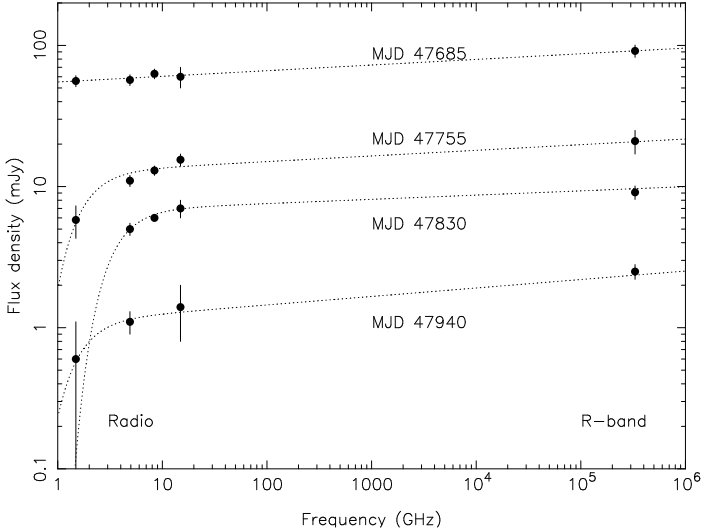}}
\caption{Evolution of the radio -- optical spectrum of GS 2023+338 /
V404 during the decline from outburst. The dotted lines are fits to
foreground thermal free-free absorption, as outlined in Appendix A.}
\end{figure}

\subsection{Polarisation}

While there have been no reported measurements of linear polarisation
from Cyg X-1, Corbel et al. (2000)
report a low level of linear polarisation, of order 2\%, from GX 339-4
in the Low/Hard X-ray state. The polarisation angle in this system has
remained approximately constant (to within uncertainties of order
10$^{\circ}$) over a period of $\geq 2$ yr, implying a fixed
orientation of the magnetic field in this system (whether parallel or
perpendicular to the jet axis is at present uncertain).  Presumably
this fixed axis corresponds to the rotation axis of the black hole
and/or inner accretion disc (the latter is in fact probably slaved to
the former).  Han \& Hjellming (1992) report a comparable level of
linear polarisation from GS 2023+338 with a constant position angle
over a period of $\sim 2$ months, again implying a relatively stable
and fixed-orientation magnetic field structure within the jet.  


\subsection{Variability timescales}

Rapid variability has been observed from the flat spectral component
in Cyg X-1 (Brocksopp et al. 1999), GX 339-4 (Corbel et al. 2000), GS
2023+338 (Han \& Hjellming 1992) and GRO J0422+32 (Shrader et
al. 1994). Furthermore, there was a clear correlation between X-ray
and radio emission in GS 2023+338 at least, implying an ongoing
coupling between the accreting and outflowing matter. In this source
the significant variability was observed on timescales of minutes,
implying size scales of order $10^{12}$ cm ($< 10^{6}$ Schwarzschild
radii for a $10$M$_{\odot}$ black hole).

In contrast some of 
the optically thin outbursts of other sources 
can be almost perfectly described by smooth power-law decays and
show little rapid variability. Hjellming \& Han (1995) show six
examples of fits to optically thin events with simple `synchrotron
bubble' models. In such cases the decay of the radio emission arises
due to adiabatic expansion losses, and so weaker radio emission is
associated with increasingly large physical structures which display
less and less variability on short timescales. Such events are also
physically decoupled and observationally uncorrelated with the
accretion disc emission following the dramatic events of the outburst.
For example, 10 days after ejection, material travelling at $\sim
0.9c$ will be $> 10^{16}$ cm ($\equiv 1500$ A.U. $\equiv 10^{10}$
Schwarzschild radii) from the central black hole. The picture is not
quite so simple however, as several transients whose radio emission is
dominated by optically thin ejections sometimes briefly display
inverted spectra (Kuulkers et al. 1999). This is interpreted as the
ejection of new components which are initially optically thick at
radio wavelengths, but which rapidly become optically thin.
Such multiple ejections of discrete blobs
has been directly observed in e.g. GRS 1915+105
(Mirabel \& Rodr\'\i guez 1994; Fender et al. 1999a).

\section{The nature of the flat spectral component}

It has long been accepted that radio emission from X-ray binaries is
synchrotron emission from material ejected from the system (Hjellming
\& Han 1995 and references therein). Such ejections, at relativistic
bulk velocities, have been directly observed in several cases
(Hjellming \& Han 1995; Mirabel \& Rodr\'\i guez 1999; Fender 2000 and
references therein). However, a simple homogenous
synchrotron--emitting source should exhibit a two component spectrum
with a spectral index of +2.5 below some frequency at which
self-absorption becomes significant, and $\alpha$ above this
frequency (with a turnover within one decade in frequency).  For a
power-law distribution of electrons of the form $N(E) dE \propto
E^{-p} dE$, this optically thin spectral index is $\alpha = (1-p)/2$.
Observed optically thin spectral indices are typically in the range
$-1 \la \alpha \la -0.5$, corresponding to electron energy indices of
$2 \la p \la 3$.  This range is approximately consistent with both simple theories
of particle acceleration and the observed cosmic-ray energy index
(Blandford \& Eichler 1987).

The radio spectra observed from these black hole systems in the
Low/Hard state is however quite different, showing a flat spectrum
which probably extends to very high frequencies.  Blandford \& K\"onigl
(1979) showed, in a model developed for AGN, that a simple
`isothermal' conical jet can produce a flat spectrum even with an
electron distribution $N(E) dE \propto E^{-2} dE$. Reynolds (1982)
explored in more detail the observed spectra from winds and jets with
a variety of geometries, magnetic fields and energetics (see also
Cawthorne 1991 for a review). Hjellming \& Johnston (1988) and Falcke
\& Biermann (1996, 1999) have discussed the application of such 
models to X-ray binaries.  Given that we recover the negative spectral
index during major optically thin outbursts, some form of the partially
self-absorbed conical jet model does seem the most likely origin of
the flat spectral component in X-ray binaries also.  In the simplest
case, a flat spectrum implies a characteristic size scale at any
frequency which is proportional to $\nu^{-1}$. Thus if the flat
spectra extend from radio to near-infrared or optical frequencies,
this implies that the jet must be self-similar over the same range in
physical size, ie. $\ga 5$ orders of magnitude.
In Fender et al. (2000) it was noted that the flat spectral component
observed from Cyg X-1 was much flatter than that observed from `flat
spectrum' AGN. This is primarily due to the high-frequency turnover,
which occurs around the millimetre band for AGN (Bloom et al. 1994),
not being observed for Cyg X-1 (or any other X-ray binary to date). As
the highest-frequency emission will arise from the smallest physical
scales for a conical jet or similar model, this lack of observed
turnover in X-ray binaries is probably due to higher densities and/or
magnetic fields in the accretion flow around a $\sim 10$M$_{\odot}$
black hole as compared to the $\geq 10^{6}$M$_{\odot}$ black holes in
AGN. If the turnover for the X-ray binaries transpires to be around
the optical or near-infrared bands, this would imply an approximate
empirical scaling of the high frequency cut-off, 
$\nu_{\rm HIGH} \propto M_{\rm BH}^{1/2}$.

Earlier interpretations of the flat radio spectra were very closely
related to these conical jet models.  The flat spectral components
observed from GS 2023+338 and GRO J0422+32 have been referred to as
`second stage' radio emission (Hjellming \& Han 1995), and it was
suggested that they originate in a wind from the accretion disc
through which the observer sees to different depths as a function of
frequency (Hjellming \& Han 1995). Furthermore it was suggested that
the slowly--decreasing flux density was as a result of a decreasing
physical size scale. Both a spherical wind and conical jet will have
an electron density which falls as $r^{-2}$ (for no pair processes),
and will produce analogous spectra (Reynolds 1982). However, the
direct observations of apparently collimated jets from Cyg X-1, 1E
1740.7-2942 and GRS 1758-258 provide strong evidence for a
collimated geometry. Furthermore, the observed linear polarisation
would not arise in a spherically symmetric source.


\section{Jet Power}

The radiative luminosity is the only quantity we can directly measure
from the flat spectral component. As non-radiative (i.e. adiabatic
expansion) losses are likely to dominate, this will be a very
conservative lower limit on the power into the jet. The observed
radiative luminosity of Cyg X-1 is $\sim 2 \times 10^{30}$ erg
s$^{-1}$ when observed up to 15 GHz in the radio band (for a distance
of $\sim 2$ kpc).  However, as argued above, there is evidence that
the flat spectral component extends across the millimetre and infrared
bands to the near infrared or even optical bands. The same flat
spectral component extending to the optical V-band will have a
radiative luminosity of $\sim 7 \times 10^{34}$ erg s$^{-1}$.

The ratio of observed (radiative) power, $\eta$, to total internal jet power is
likely to be $\leq 5$\%, a figure based both on theory (Blandford \&
K\"onigl 1979) and on observations of GRS 1915+105 (Fender \& Pooley
2000). This is because the electrons lose energy primarily as a result
of adiabatic expansion, and not via synchrotron or inverse Compton
processes (although above some cut-off frequency these radiative
losses will dominate).

The power in such a jet can be estimated as

\[
L_{\rm jet} \ga 10^{36} 
\left(\frac{d}{2{\rm kpc}}\right)^2 
\left(\frac{\nu_{\rm HIGH}}{10^{14}{\rm Hz}}\right)
\left(\frac{S_{\nu}}{15{\rm mJy}}\right)
\left(\frac{\eta}{0.05}\right)^{-1} 
F(\Gamma,i)
\]

erg s$^{-1}$ where $\nu$ is the high-frequency cut-off to the flat spectral
component, $S_{\nu}$ is the flux density (dereddened, if necessary)
measured at that frequency, and $F$($\Gamma$, i) is an approximate
correction factor for relativistic bulk motion (see below).
The parameters have been scaled for Cyg
X-1, with a high-frequency cut off around 3$\mu$m.

The observed X-ray luminosity of Cyg X-1 in the Low/Hard state is $\sim 3
\times 10^{37}$ erg s$^{-1}$ (Nowak et al. 1999; di Salvo et
al. 2001). So, based on only the following assumptions:

\begin{enumerate}
\item{The flat spectral component extends to near-infrared / optical
bands}
\item{The radiative efficiency of the jet is $\leq 5$\%}
\item{Relativistic beaming of the radio emission is not significant}
\end{enumerate}

it is concluded that
{\em the power in the jet, in the Low/Hard X-ray state, is $\geq 5$\% of
the total accretion luminosity as observed in X-rays.}

However, the third assumption is certainly worth considering more
carefully, since observations of other X-ray binaries certainly have
revealed evidence for outflows at relativistic velocities. The
luminosity based on the above inequality can be significantly affected
by bulk relativistic motions which will (a) Doppler shift the observed
frequencies, (b) `Doppler boost' the observed flux densities
(aberration), and (c) add a significant amount of bulk kinetic energy
to the power requirements.

For a flat ($\alpha=0$) spectral component of flux density $S_{\nu}$
extending to frequency
$\nu$, $L \propto \nu S_{\nu}$. For bulk motion at velocity $\beta c$, 
the Lorentz factor is 
$\Gamma = (1-\beta^2)^{-1/2}$ and corresponding relativistic Doppler
factor $\delta = [\gamma(1-\beta \cos \theta)]^{-1}$.
The transformation from the comoving
to observer's frames is $\nu_{\rm obs} = \delta \nu_0$ and
$S_{{\nu}{\rm obs}} =
\delta^{k-\alpha} S_0$, where $k=3$ for a single discrete component,
and $k=2$ for a simple jet emitting isotropically in its rest frame
(Rybicki \& Lightman 1979; Cawthorne 1991); therefore $L_{\rm obs} = L_0
\delta^{1+k-\alpha}$. Assuming $\alpha=0$ and $k=2$ (observationally
$k$ is found to be nearer to 2 than 3 for GRS 1915+105 -- Mirabel \&
Rodr\'\i guez 1994; Fender et al. 1999a), $L_0 = L_{\rm obs}
\delta^{-3}$. In addition, taking into account bulk relativistic
motion, the total jet power must be multliplied by $\Gamma$. So,
compensating for both Doppler shifts and kinetic energy, the
relativistic `correction factor' $F(\Gamma,i) = \Gamma \delta^{-3}$.
This function is plotted for a range of velocities, at all
inclinations, in Fig 6.  Note most importantly that as the effect
becomes more significant, ie. for the highest velocities, the chances of
{\em overestimating} the jet power from observations (ie. when $F(i) <
1$) decrease rapidly; for a random sample of inclination angles the
total jet power is more likely to be {\em underestimated} if
relativistic bulk motion
is neglected.  This is because to first order the radiation
will be beamed within an angle $\sim 1/\Gamma$ of the forward
direction of motion of the ejecta.  In the case of Cyg X-1, best
estimates of $i$ (presumed to be equal to the orbital inclination) are
around 30$^{\circ}$; inspection of Fig 6 shows that at this
inclination the above method will not {\em overestimate} the total jet
power by more than a factor of 3, and for high velocities
(e.g. $v=0.98c$, $\Gamma=5$) we may be {\em underestimating} the jet
power by up to a factor of 5. For example, for Cyg X-1, assuming
$\nu_{\rm HIGH} = 10^{14}$ Hz, $\eta=0.05$, $\Gamma=5$ and
$i=30^{\circ}$, $L_{\rm jet} / L_{\rm X} \geq 0.2$.

Furthermore, Brocksopp et al. (1999) and Corbel et al. (2000) report
approximately linear relations between the X-ray and radio fluxes of
Cyg X-1 and GX 339-4 when these sources are in the Low/Hard X-ray
state, a trend also observed in GS 2023+338 (Han \& Hjellming
1992). If this relation holds for all sources in the Low/Hard X-ray
state then we can estimate the radio flux density based on X-ray
observations, and vice-versa. Empirically, the relation is

\[
S_{\nu,{\rm cm}} \sim 75 \left(\frac{S_X}{{\rm Crab}}\right)
\phantom{000} {\rm mJy}
\]

Note
that since there is evidence for inversion of the spectrum as the
source weakens, this correlation will be strongest for higher radio
frequencies.  Note also that this relation is approximately equivalent
to 1 mJy of radio flux density for 1 RXTE ASM ct/sec; the quantitative
nature of this relation will be explored elsewhere.  This
approximately linear relation indicates that whatever the previous
accretion and ejection history, when in the Low/Hard X-ray state, black
hole X-ray binaries produce a jet whose power is an approximately
fixed, and probably large, fraction of the accretion
luminosity. Furthermore, the comparable radio luminosity of the four
persistent Low/Hard state sources (Fig 5; see also Fender \& Hendry
2000) agrees with this interpretation as long as the X-ray
luminosities of the four sources are within an order of magnitude of
each other, which seems to be the case.

\begin{figure}
\centerline{\epsfig{file=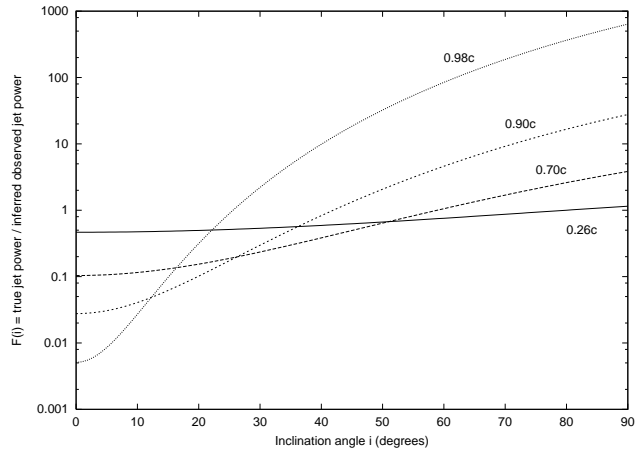,width=6cm,angle=270}}
\caption{Correction factor $F(i)$ for estimating jet power, as a
function of angle $i$ to the line of sight, for a range of intrinsic
bulk velocities. Note that the higher the velocity, the more likely it
is that the jet power will be {\em underestimated} if Doppler boosting 
is neglected.}
\end{figure}

\section{Low/Hard states in GRS 1915+105}

GRS 1915+105 is a very luminous X-ray binary black hole candidate and
the first Galactic source of superluminal radio jets (Mirabel \&
Rodr\'\i guez 1994; Fender et al. 1999a; Belloni et al. 2000). Its
relevance to this discussion is twofold : 

\begin{enumerate}
\item{It was the first X-ray binary source for which there was firm
evidence for an extension of a flat-spectrum synchrotron component 
extending through the millimetre (Ogley et al. 2000, Fender \& Pooley 2000)
to the near-infrared (Fender et al. 1997b; Mirabel et al. 1998;
Fender \& Pooley 1998, 2000) bands. Most of this evidence is in the form of
oscillation events, associated with repeated accretion cycles, which
are identical at radio, mm and near-infrared wavelengths.}
\item{The source occasionally displays `plateau' X-ray states, which
are reminiscent of the canonical Low/Hard state, but more luminous
(Belloni et al. 2000). Associated with these states is a steady,
flat-spectrum synchrotron component in the radio band (Fender et al. 1999a).}
\end{enumerate}

The plateau states (class $\chi$, state C of Belloni et al. 2000)
are dominated by a power-law component, with little
or no disc contribution in the X-ray band (Belloni et al. 2000). In
this sense they are similar to the Low/Hard X-ray state. Fig 7
illustrates the radio -- infrared spectrum of the source during one
such state, in 1996 August. Other examples of the plateau also show a
flat spectrum in the range 1--15 GHz (Hannikainen et al. 1998a; Fender
et al. 1999a), but the 1996 August event appears to be the only plateau
with simultaneous infrared observations (Bandyopadhyay et al. 1998).
So again, in circumstances which are rather different (a more luminous,
exotic source) from the `quiet' Low/Hard state of e.g. Cyg X-1, we
again find a flat-spectrum, with evidence for absorption at low radio
frequencies and a high-frequency extension to the near-infrared. The
reader is reminded that the synchrotron oscillation events in GRS
1915+105 are also associated with brief excursions (or `dips') to the
same state which characterises the `plateaux' and, in these cases
there is little doubt that the synchrotron spectrum does extend to the
near-infrared (Fender et al. 1997c; Pooley \& Fender 1997; Eikenberry
et al. 1998, 2000; Mirabel et al. 1998; Fender \& Pooley 1998).
Very recently Dhawan, Mirabel \& Rodr\'\i guez (2000) have confirmed that the
plateau state in GRS 1915+105 is associated with a compact jet.

\begin{figure}
\centerline{\epsfig{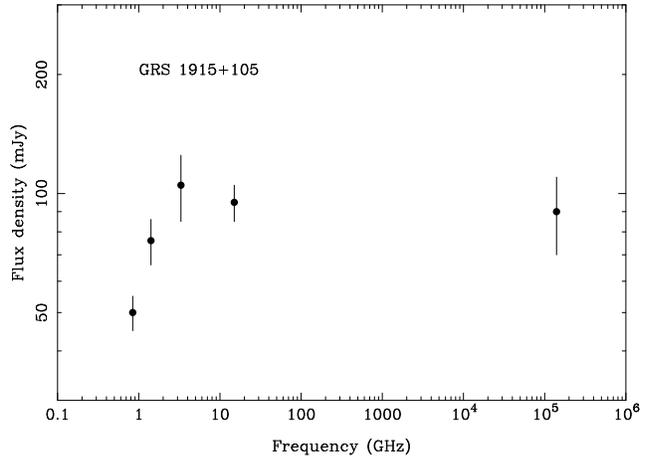}}
\caption{The radio--infrared spectrum of GRS 1915+105 during the
`plateau' state in August 1996 ($\sim$ MJD 50300). Data are from
Pooley \& Fender (1997), Hannikainen et al. (1998a), Bandyopadhyay et
al. (1998).}
\end{figure}

\section{Conclusions}

There are four persistently accreting black hole candidates in our
Galaxy which spend most of their time in the power-law-dominated
Low/Hard X-ray state. These are Cyg X-1, GX 339-4, 1E1740.7-2942 and
GRS 1758-258. It is shown that all four sources have comparable radio
spectra, namely weak emission with a flat/inverted ($0 \la \alpha \la
0.5$) spectrum. Furthermore in two, probably three, of the four systems,
jets have been directly imaged at radio wavelengths.

It is unusual for transient black hole X-ray sources to spend an
extended period in the Low/Hard X-ray state, but this was the case for
at least three systems, GS 2023+338, GRO J0422+32 and GS 1354-64,
during periods of extensive radio coverage. All three systems were
found to have radio spectra which, after initial optically thin events 
associated with the state transition, 
 were very similar to those of the
persistent sources in the Low/Hard state, and very different from the
bright, optically thin outbursts seen from most other transients 
(Fig 4). As previously suspected (Fender 2000), all black hole X-ray
binaries, whether persistent or transient, display a flat radio
spectrum when in the Low/Hard X-ray state. For further comparison of
the observational properties of Low/Hard state transients, see
Brocksopp et al. (2001).

The flat radio spectra observed from these sources do not show
high-frequency cut-offs, but on the contrary show evidence for
extension into the millimetre, infrared and maybe even optical
bands. All three transient sources show some correlated radio :
optical behaviour, and have approximately flat spectra from the radio
through to the optical bands. All three systems also show evidence for
absorption at lower frequencies as the flux levels decline, possibly
(although certainly not conclusively) evidence for increasing
free-free absorption as the jet shrinks back towards the core of the
system. A similar radio--millimetre--infrared spectrum is observed
from the jet source GRS 1915+105 when in a spectrally similar
(ie. dominated by a power-law), but more luminous hard X-ray state
(Fig 7).

In the case that the flat spectral component {\em does} extend to the
optical or near-infrared bands, and the jet has a low ($\la 5$\%)
radiative efficiency, then the luminosity of the jet is likely to
exceed 5\% of the observed X-ray luminosity, and is furthermore likely
to scale linearly with the X-ray flux. All this evidence supports a
jet--disc `symbiosis' model (Falcke \& Biermann 1996, 1999) in which
the jet luminosity is some fixed, and probably large, fraction of the
accretion luminosity. The observed comparable
correlation between the radio (jet)
and hard X-ray (Comptonised) emission in different systems
implies that these two
components do not have very different beaming; this in turn implies
that unless the hard X-ray emission is strongly beamed (possibly via
synchrotron self-Compton in the jet) then the jet in the Low/Hard
state is unlikely to have a very large Lorentz factor ($\Gamma \la 5$
or so).

Further observations to determine the high-frequency extent of the
flat spectral component are of great importance. Confirmation of a
high fraction of the accretion luminosity being diverted into the jet
during hard, power-law-dominated X-ray spectral states will have great
significance for our understanding of the accretion flow near a black
hole, not least for models of advection-dominated accretion.

\section*{Acknowledgements}

I would like to thank many people including Stephane Corbel, Chris
Done, Heino Falcke, Bob Hjellming, Kinwah Wu, Catherine Brocksopp,
Mariano Mendez, Chris Shrader and Michiel van der Klis for useful
discussions and information. In addition I would like to thank Luis
Rodr\'\i guez and Felix Mirabel for supplying the maps presented in Fig 2.
Finally I would like to thank the referee, Ralph Spencer, for many
in-depth comments which helped to improve the paper. This research has
made use of the SIMBAD database, operated at CDS, Strasbourg, France.
RPF was supported during the initial period of this research by EC
Marie Curie Fellowship ERBFMBICT 972436.

\appendix
\section{Foreground free-free absorption of a compact jet}

In order to see if the radio--optical spectra can be fit by
a single power-law which is absorbed by free-free emission at longer
radio wavelengths, the four spectra in Fig 5 have been fit by the
following simple model

\[
S_{\nu} = S_0 \nu^{\alpha} e^{-\tau_1 \nu^{-2.1}}
\]

where $S_0$ is the amplitude of the power law, $\alpha$ is the
unabsorbed spectral index and $\tau_1$ is the optical depth at 1 GHz
(free-free optical depth $\propto \nu^{-2.1}$ at radio wavelengths,
e.g. Gordon 1988). The fits are illustrated in Fig 5 and tabulated in
Table A1. It is interesting that the fits give an unabsorbed spectral
index which is approximately constant at a slightly inverted value.
The amplitude decreases and the optical depth to free-free absorption
increases for the first three epochs, then decreases again for the
final epoch. If this is a correct modelling of the data, then two
possibilities suggest themselves -- either a cloud of ejecta from the
outburst slowly moved outward and enveloped the jet and then
eventually dispersed, or as the jet weakened its characteristic size
scale decreased and it withdrew back within a shroud of material still
enveloping the system post-outburst. Note that in the very inner
regions of the SS 433 jet the radio spectrum becomes inverted,
probably due to free-free absorption (Paragi et al. 1999).

\begin{table}
\caption{Fits to the radio--optical spectra presented in Fig 7 to the
function $S_0 \nu^{\alpha} e^{-\tau_1 \nu^{-2.1}}$}
\center
\begin{tabular}{cccc}
MJD & $S_0$ & $\alpha$ & $\tau_1$ \\
\hline
47685 & $55.1 \pm 1.8$ & $0.04 \pm 0.00$ & $0.00 \pm 0.13$ \\
47755 & $12.5 \pm 1.1$ & $0.04 \pm 0.01$ & $1.91 \pm 0.71$ \\
47830 & $6.6 \pm 0.4$ & $0.03 \pm 0.01$ & $9.29 \pm 2.51$ \\
47940 & $1.1 \pm 0.1$ & $0.06 \pm 0.01$ & $1.56 \pm 0.47$ \\
\hline
\end{tabular}
\end{table}

Could the low-frequency inversion be due instead to synchrotron
self-absorption ? This seems hard to reconcile with the observations
since, if the emission does arise in a conical jet, then the lowest
frequencies, which we observe to be increasingly suppressed, should
arise on the largest scales. Thus in order for synchrotron
self-absorption to affect lowest frequencies, it would appear that we
would require some enhancement of magnetic field and/or particle
number density to occur towards the end of the jet. Perhaps as the
stratified jet recedes towards its base this could happen; without
entering into modelling of the jet itself this is unclear. Further
detailed low-frequency observations of the jet spectrum are desirable
to investigate this.

\end{document}